# Open GENIE – Analysis and Control.

S. I. Campbell, F. A. Akeroyd, C. M. Moreton-Smith

*ISIS Facility, CLRC Rutherford Appleton Laboratory,
Oxfordshire, OX11 0QX, UK.*

*Abstract*

Open GENIE is the successor to the widely used and high successful GENIE program that is used as the principal front-end neutron data analysis tool at the ISIS spallation neutron source. In this paper, the background and motivation for the 'Open GENIE' project are briefly reviewed, as is the progress to date. Some of the ideas and concepts making Open GENIE unique are demonstrated. Emphasis is given to how Open GENIE is being used as an integral component of the new ISIS control system and how it is expected to develop in the future and how they will impact the field of neutron scattering data analysis is also discussed.

*Introduction*

The original GENIE program was developed in the 1980s when the ISIS pulsed spallation neutron source was first commissioned. For the past decade, a stable OpenVMS (AXP/VAX) version of the software (GENIE-V2) [1] has been in use at *ISIS* (Rutherford Appleton Laboratory, Oxfordshire, UK), and at several scientific and commercial establishments world-wide. Constant improvements in the capabilities of the neutron scattering instruments at ISIS, and in the hardware and software available for data analysis, has necessitated a re-write of GENIE.

The new GENIE is known as Open GENIE to reflect the intention that the software be used on a wide variety of different computers and operating systems. It is aimed at supplying scientists with an inexpensive computer package which provides them with access to their experimental data, so that they may analyse the data as required and display the results in a useful format, usually in a graphical form.

*Overview*

This paper is not intended as an extensive guide to the features of Open GENIE or indeed a replacement for the user [2] or reference [3] manuals, it will instead focus on the basic use of Open GENIE as a data reading, analysis and visualisation tool together with some specific features that have been incorporated to enable Open GENIE to be used provide scripting control to ISIS instruments. The list of operating systems that Open GENIE supports is ever growing and the platforms that stable versions are available for include Microsoft Windows XP/2000/NT/9x, UNIX (in many of its flavours), Compaq OpenVMS and Mac OS X. Open GENIE has reached version 2.2 which is available for download (or installation on windows directly from the web) from the ISIS website [4] and is distributed under the GNU Public License (GPL).

When you run Open GENIE you have a choice over which kind of interface you would like to use, namely either a command line (default) or a Tk/Tcl GUI interface (see Figure 1).

### *The Three Wishes*
One of the most useful ways in which to view the functionality of Open GENIE is to take the point of view of someone who is doing an experiment and wants to use Open GENIE to analyse their data. Whether they are directly aware of this or not they will need *Access* to their data, tools for *Manipulating* their data and a means of *Visualising* their data. The rest of this paper discusses how Open GENIE achieves these basic needs elegantly and efficiently.

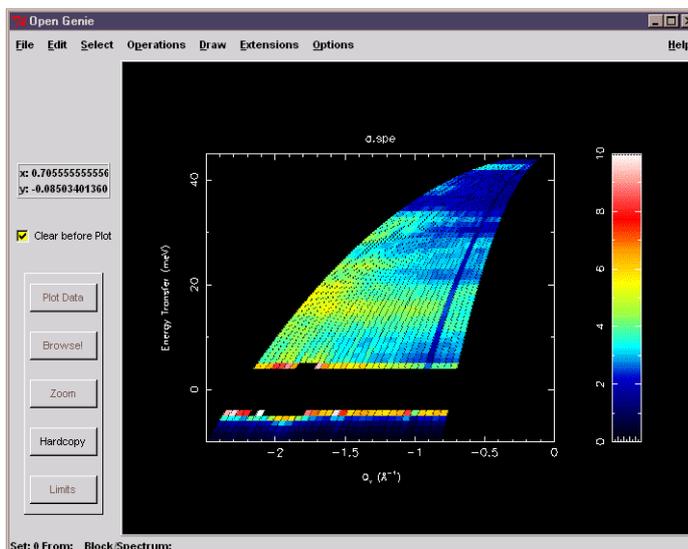

Figure 1: The Open GENIE graphical user interface.

### *The first wish – Access the data*
There are only two things that a user of a data analysis program must specify to obtain data; the source of the data and the item to read. Open GENIE transparently detects different data formats and performs automatic conversions when an item is read from a particular source. As an example, if the source is a NeXus [5] file and the item being read is the first NXentry, the conversion will be into an internal one or two-dimensional Open GENIE workspace. The command required to perform this operation is simply;

```
>> mydata = get(1, "mydatafile.nxs")
```

Whereas the command to read the first spectrum in an ISIS raw data file is;

```
>> mydata = get(1, "sic12345.raw")
```

Open GENIE also supports flexible reading from ASCII files (via the *ASCIIFILE* command).

### *The second wish – Manipulate the data*
Open GENIE allows basic arithmetic and simple operations on real and integer variables. As well as providing operations on these basic data types, the object based architecture requires that these operations extend logically, first to operate on arrays of data (integer or real) and second to operate on workspaces (like C structures) as heterogeneous collections of different types of data.

In Open GENIE, the only distinction between functions and procedures is that functions must return a result. As an example, a functional expression may be used to do a monitor efficiency correction.

```
>> mon = mon / (unt * (1- exp( -8.3 * sig * mon.x )))
```

In this example all the different data types are combined by a clear set of operational rules. "mon" is a complete spectrum workspace with structure members (fields) containing defined units, X, Y and Error arrays. The denominator of the expression is an array expression based on the X array from the original workspace; the array is multiplied, the exponential taken and subtracted as a whole. Because the numerator is a workspace, specific (customisable) conversion rules are called to specify how the division actually takes place. By default the Y array member/field is used if it is present. All calculations on Open GENIE workspaces automatically use Gaussian errors but once again this behaviour is customisable by the user, so a different error scheme or multiple schemes may be chosen.

Open GENIE provides a rich procedural language (GCL) to allow the control of specific analysis algorithms, and to allow new functions and commands to be defined (: is used rather than / for a qualifier in a function). An example procedure for normalising data is given below.

```
PROCEDURE Norm_sum
PARAMETERS first last
RESULT gspec
LOCAL mon
LOCAL unt=3.584e-3 sig=0.013

    mon = s(1)
    gspec = focus:t(first:last)/(last-first+1)
    mon = mon/(unt*(1-exp(-8.3*sig*mon.x)))
    gspec = rebin(gspec, mon.x)
    units/lam gspec

ENDPROCEDURE
```

Now the procedure can be used just like any other Open GENIE function.

```
>> set/file "iris$disk0:[irsmgr.data]irs09089.raw"
>> final = Norm_sum(2,20)
```

Due to the structure of Open GENIE (Figure 2), it is possible to integrate a number of analysis routines either directly in GCL (as in the above example) or in a Module (user created C or FORTRAN dynamic/shared library).

### *The third wish – Visualise the data*
The final requirement of data visualisation is in principle a fairly simple one. Some very sophisticated tools exist for real time visualisation of multi-dimensional data and it is not the purpose of Open GENIE to duplicate the functionality of a product such as IDL, MATLAB or even a technical plotting package like SigmaPlot or Origin. The aim within Open GENIE is to allow all classes of data which can be manipulated to be displayed efficiently, easily and in a way appropriate to the analysis. For further details of how to display data using external packages such as IDL, MatLAB, SigmaPlot or Excel please refer to [6].

## *How is Open GENIE Organised?*

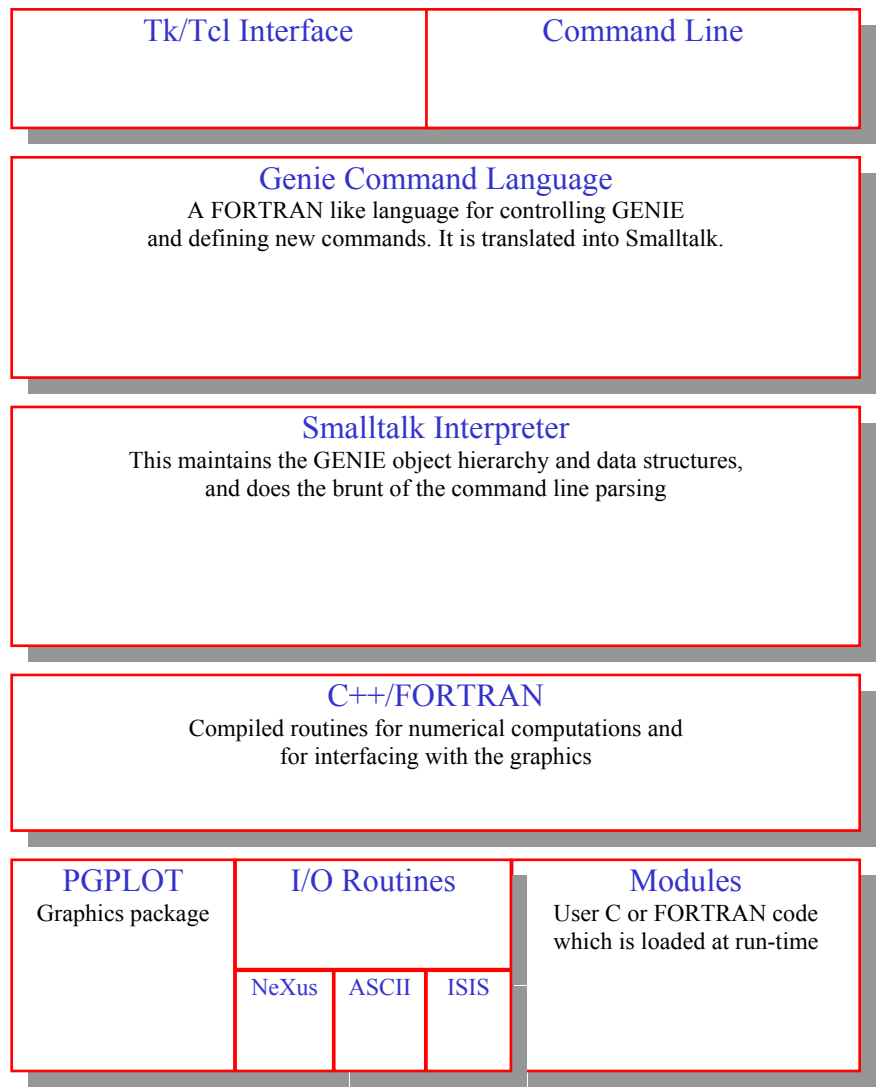

**Figure 2: Structure of Open GENIE.**

To make for easy usage, Open GENIE provides procedures which emulate the basic *Display*, *Plot* and *Multiplot* graphics commands as in the previous version of GENIE. These commands automatically handle data based on a multi-dimensional data (see Figure 3) and use PGPLOT [7]. For flexibility, the high level graphics procedures are written from a basic set of graphics primitive functions. These allow the generic procedures to be modified or to be replaced with display procedures suited to a particular task. For further details on these routines, please refer to the Open GENIE Reference Manual [3].

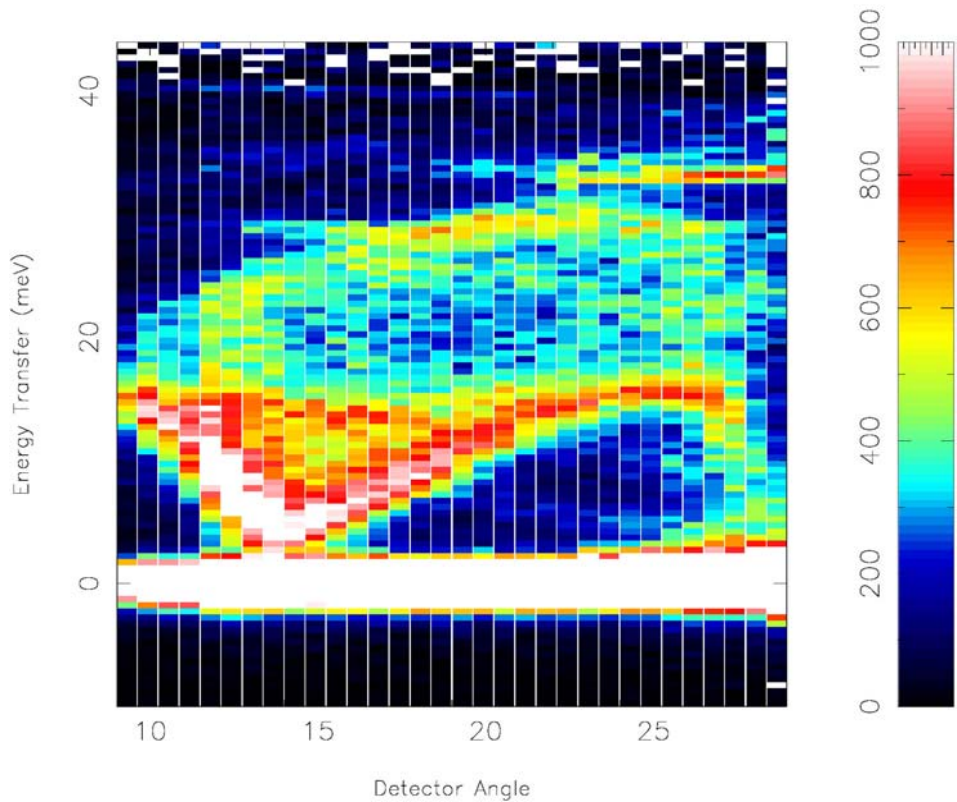

**Figure 3: Open GENIE plot of multi-dimensional data.**

### *DCOM and Scripting*

Traditionally instrument control at ISIS has been performed via an AlphaServer (or Station) running OpenVMS. Recently, a decision has been made to migrate to a system based on LabVIEW [8] running on Microsoft Windows. At the moment, experiment control scripts are written in OpenVMS DCL scripting language which of course isn't available on a Windows platform. After some investigation and research into what users wanted from a scripting language it was decided to use Open GENIE's own command language (GCL).

The details on the configuration of the new ISIS control system can be found in a paper by Akeroyd et al. in this volume [9]. In order for the scripting language to interact with the instrument control program (Ray of Light) which uses LabVIEW, it was necessary to incorporate DCOM support into Open GENIE. This enables Open GENIE to communicate with any program that implements a DCOM (or ActiveX) automation interface and also to be called by any ActiveX program (e.g. Visual Basic). For example, if you wanted to transfer some data into Microsoft Excel then this can be achieved by using the following code;

```
PROCEDURE dcom_excel_test
LOCAL e excel workbooks workbook sheet range
    excel=dcom:create("Excel.Application")
    # set property to make Excel visible
    dcom/call/propput excel "Visible" "" 1
    # read an Excel property
    workbooks=dcom:call:propget(excel,"Workbooks")
    workbook=dcom:call:propget(workbooks,"Add")
    sheet=dcom:call:propget(excel,"ActiveSheet")
```

```
        range=dcom:call:propget(sheet,"Range","","A1:E1")
        # create an array of values to send to Excel
        e=dimensions(5)
        fill e 1.0 1.0
        dcom/call/propput range "Value" "" e
        # avoid being prompted to save workbook on exit
        dcom/call/propput workbook "Saved" "" 1
ENDPROCEDURE
```

The scripting syntax is based on the existing set of instrument control commands but slightly modified to improve their functionality and flexibility. As the scripts are executed within an Open GENIE session where we also have access to all of the data access and reduction/analysis routines, we can start to build in some intelligence into our instrument control scripts (i.e. on-the-fly analysis). For example, if we wanted to make an experimental measurement until we could fit a particular peak (or collection of peaks) to a given model within a stated accuracy then this would now be quite easy to achieve. The following code is a simplified example of how we might achieve this;

```
# Let us first make the default input
# source the data acquisition system.
ASSIGN $dae
# Begin the run
BEGIN

LOOP
    data2fit = get(10)   # Read in the data to fit
    fiterror = myfitting_routine(data2fit)
    EXITIF (fiterror < 0.1)
ENDLOOP

# Stop the measurement
END
```

The scripting framework that has been developed makes it very easy to set or read any variable that exists within the LabVIEW control program. Each quantity in LabVIEW is described by its VI (virtual instrument) name, the label of the value to set (i.e. the control value) and the label for the read back value. These are then assigned to a short block name (e.g. temp1, field, phase, pressure, etc…) which can be set by using the *CSET* command (e.g. `CSET temp1=100.0`) or read by using the *CSHOW* command (e.g. `CSHOW temp1`). One important aspect of the project is to develop a scripting interface was the ability to control the instrument from a remote computer. This goal has been achieved, it is now possible to control the instrument from an instance of Open GENIE running anywhere on the network (presuming that the user has the necessary user rights). Work has currently begun to provide a remote graphical dashboard interface to control and/or monitor the instrument. For more details on the scripting interface please refer to [10].

*Using MATLAB to access and display*
The following is a brief example of how to access and view (Figure 4) experimental data using from within MATLAB using Open GENIE's COM Automation interface.

```
og = actxserver('OpenGENIE.Application');

invoke(og,'AssignHandle','w=get(3,"c:/data/irs11944.raw")','')
```

```
ans =
    -1

invoke(og,'AssignHandle','wy=w.y; wx=centre_bins(w.x);','')

ans =
    -1

wx = invoke(og, 'GetValue', 'wx');
wy = invoke(og, 'GetValue', 'wy');

plot (wx,wy)
```

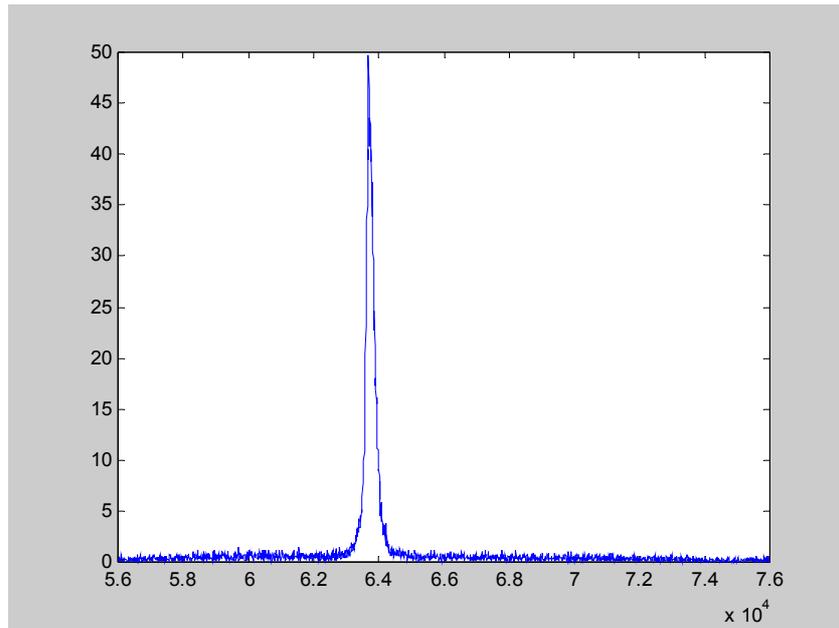

**Figure 4: MatLAB plot using Open GENIE to access data.**

Open GENIE also provides a number of interfaces to access it directly from an external user program written in a number of different languages and packages. All the routines take the form G*n*_ROUTINE_NAME, where *n* is X for access from C/C++, F for access from FORTRAN and I for access from IDL. For example, GI_ACTIVATE_SESSION will activate an Open GENIE session from within IDL. This is called the GDAI (Genie Data Access Interface), which is also available as a COM component that can be called from languages such as Visual Basic etc..

*Final Remarks*
More detailed information on the current state of the Open GENIE project may be found from the World Wide Web[4] and specific details of Open GENIE functionality may be found in the manuals [2, 3].

A brief overview of how Open GENIE is being used as a scripting tool for the new PC based instrument control system at ISIS has been presented. This demonstrates the advantage of having the power of Open GENIE coupled with instrument control routines.

*References*

[1] GENIE-II, Rutherford Appleton Laboratory Report RAL-86-102

[2] Open GENIE User Manual, Rutherford Appleton Laboratory Technical Report, RAL-TR-2000-002 (or online at http://www.isis.rl.ac.uk/GenieUserManual/ ).

[3] Open Genie Reference Manual, Rutherford Appleton Laboratory Technical Report, RAL-TR-1999-031 (or online at http://www.isis.rl.ac.uk/GenieReferenceManual/ ).

[4] Open GENIE Web Site – http://www.isis.rl.ac.uk/OpenGENIE/ .

[5] NeXus Web Site – http://www.neutron.anl.gov/nexus/ .

[6] http://www.isis.rl.ac.uk/OpenGENIE/visualisation/ .

[7] PGPLOT Web Site – http://www.astro.caltech.edu/~tjp/pgplot/ .

[8] National Instruments LabVIEW (http://www.ni.com/labview/ ).

[9] F.A. Akeroyd, S.I. Campbell, C.M. Moreton-Smith, "The New ISIS Instrument Control System", NOBUGS2002/040, cond-mat/0210468 in arXiv.

[10] http://www.isis.rl.ac.uk/computing/projects/ .